%% file: submission
\documentclass{llncs}

\usepackage{graphicx}
\usepackage{dcolumn}
\usepackage{bm}
\usepackage{bbm,dsfont}
\usepackage{a4wide}
\usepackage{amsmath,amssymb,amsfonts,latexsym,stmaryrd}

\usepackage[PostScript=dvips]{diagrams}
\usepackage{QED}
\usepackage{epsfig}
\usepackage{euscript}
\usepackage{tikz,ifdraft}
\usetikzlibrary{arrows,positioning,matrix,backgrounds,calc,fit}
\usepackage{url}

\newcommand{\plusminus}{\pm}

\newcommand{\beqa}{\begin{eqnarray*}}
\newcommand{\eeqa}{\end{eqnarray*}\par\noindent}

\newcommand{\XX}{\mathbf{X}}

\newcommand{\lrarr}{\longrightarrow}
\newcommand{\rarr}{\rightarrow}

\newcommand{\II}{\mathcal{I}}

\newcommand{\ie}{\textit{i.e.}~}

\newcommand{\EE}{\mathbf{E}}

\newcommand{\Real}{\mathbb{R}}

\newcommand{\id}{\mathsf{id}}

\newcommand{\AND}{\; \wedge \;}

\newcommand{\UU}{\mathcal{U}}

\newcommand{\XOR}{\oplus}

\newcommand{\MM}{\mathcal{M}}

\renewcommand{\emph}{\textbf}

\newcommand{\prob}{\mathsf{Prob}}

\newcommand{\xp}{x^{+}}
\newcommand{\xm}{x^{-}}

\newcommand{\Pp}{\prob^{\pm}}
\newcommand{\rpm}{r^{\plusminus}}
\newcommand{\Xpm}{X^{\plusminus}}
\newcommand{\Ypm}{Y^{\plusminus}}
\newcommand{\fpm}{f^{\plusminus}}

\newcommand{\vs}{\varsigma}
\newcommand{\IF}{\mathds{1}}
\newcommand{\spm}{s^{\vs}}
\newcommand{\mupm}{\mu^{\pm}}
\newcommand{\nupm}{\nu^{\pm}}
\newcommand{\rv}[1]{\mathbf{#1}}
\newcommand{\Seq}{\mathsf{Seq}}
\newcommand{\Ixy}{\II}

\begin{document}

\title{An Operational Interpretation of Negative Probabilities and No-Signalling Models}

\author{Samson Abramsky and Adam Brandenburger}

\institute{Department of Computer Science, University of Oxford \\ Stern School of Business and Center for Data Science, New York University\\ \ \\
\texttt{samson.abramsky@cs.ox.ac.uk} \\ \texttt{adam.brandenburger@stern.nyu.edu}}

\maketitle

\begin{abstract}
Negative probabilities have long been discussed in connection with the foundations of quantum mechanics. We have recently shown that, if signed measures are allowed on the hidden variables, the class of probability models which can be captured by local hidden-variable models are exactly the no-signalling models. However, the question remains of how negative probabilities are to be interpreted. In this paper, we present an operational interpretation of negative probabilities as arising from \emph{standard} probabilities on \emph{signed} events. This leads, by virtue of our previous result, to a systematic scheme for simulating arbitrary no-signalling models.
\end{abstract}

\section{Introduction}

Negative probabilities have been discussed in relation to quantum mechanics by many authors, including Wigner,  Dirac and Feynman \cite{wigner1932quantum,Dirac42,Feynman1987negative}. For example, Feynman writes:
\begin{quotation}
The only difference between a probabilistic classical world and the equations of the quantum world is that somehow or other it appears as if the probabilities would have to go negative \ldots
\end{quotation}

\noindent The separation of quantum from classical physical behaviour in results such as Bell's theorem \cite{bell1964einstein} is expressed in terms of \emph{local realistic models},  in which ontic (or hidden) variables control the behaviour of the system in a classical fashion, satisfying the constraints of \emph{locality} and \emph{realism}. The content of Bell's theorem is exactly that no such model can give rise to the behaviours predicted by quantum mechanics.

However,  if we allow negative probabilities on the ontic (or hidden) variables of the model, the situation changes radically.

As a warm-up example, we shall consider the following scenario due to Piponi\footnote{See the blog post at \url{http://blog.sigfpe.com/2008/04/negative-probabilities.html}}, which, while artificial, is appealingly simple, and does convey some helpful intuitions.

We shall consider a system comprising two bit registers, $A$ and $B$. We can perform the following tests or observations on these registers on each run of the system:
\begin{itemize}
\item We can read $A$ or $B$, \emph{but not both}.
\item Alternatively, we can observe the value of $A \XOR B$, the \emph{exclusive or} of $A$ and $B$.
\end{itemize}
What we find by empirical observation of the system is that, in every run:
\begin{enumerate}
\item When we read $A$, we always get the value 1.
\item When we read $B$, we always get the value 1.
\item When we observe $A \XOR B$, we always get the value 1.
\end{enumerate}
From 1 and 2, we infer that $A =1$ and $B=1$, but this contradicts our observation that $A \XOR B = 1$.

We can try to explain this apparently contradictory behaviour as follows:
\begin{itemize}
\item On each run of the system, the registers are set to one of four possible combinations of values:
\begin{center}
\begin{tabular}{c|c|c|c|c} \hline
$AB$ & $00$ & $01$ & $10$ & $11$ \\ \hline
\end{tabular}
\end{center}
\item This joint value is sampled from a probability distribution:
\begin{center}
\begin{tabular}{c|c|c|c|c} \hline
$AB$ & $00$ & $01$ & $10$ & $11$ \\ \hline
& $p_1$ & $p_2$ & $p_3$ & $p_4$ \\ \hline
\end{tabular}
\end{center}
\end{itemize}
It is easily seen that no such distribution, where $p_i \geq 0$, $i=1,\ldots , 4$ and $\sum_i p_i = 1$, can yield the observed behaviour of the system.
However, consider the following \emph{signed distribution}:
\[ \begin{array}{lcr}
p(00) & = & -1/2 \\
p(01) & = & 1/2 \\
p(10) & = & 1/2 \\
p(11) & = & 1/2
\end{array}
\]
Note that the probability of reading the value 1 for $A$ is
\[ p(A=1) \; = \; p(10) + p(11) \; = \; 1 , \]
and similarly
\[ p(B=1) \; = \; p(01) + p(11) \; = \; 1 . \]
Also,
\[ p(A \XOR B = 1) \; = \; p(10) + p(01) \; = \; 1. \]
Finally, the measure is normalised:
\[ p(00) + p(01) + p(10) + p(11) \; = \; 1 . \]
Also, note that the negative value $p(00) = - 1/2$ can never be observed, since we cannot read the values of both $A$ and $B$ in any given run.
In fact, the only events which are accessible by direct observation are the following:
\[ \begin{array}{lcl}
A = 1 & \quad & \{ 10, 11 \} \\
A = 0 & \quad & \{ 00, 01 \} \\
B = 1 & \quad & \{ 01, 11 \} \\
B = 0 & \quad & \{ 00, 10 \} \\
A \XOR B = 1 & \quad & \{ 10, 01 \} \\
A \XOR B = 0 & \quad & \{ 00, 11 \}
\end{array}
\]
All of these events get well-defined, non-negative probabilities under our signed distribution, and moreover the complementary pairs of events corresponding to the different possible outcomes for each run, given that a particular choice of quantity to test has been made, yield well-defined probability distributions:
\[ p(A = 1) + p(A = 0) \; = \; 1, \quad p(B= 1) + p(B = 0) \; = \; 1, \quad p(A \XOR B = 1) + p(A \XOR B = 0) \; = \; 1. \]
Of course, unless we can give some coherent account of the negative value appearing in the signed measure, it could be said that we have simply explained one mystery in terms of another.

Before addressing this point, we shall turn to a more substantial example, which plays a central r\^ole in much current work in quantum information and foundations.

We shall now consider a scenario where Alice and Bob each have a choice of two 1-bit registers (or  ``measurement settings''); say $a$ or $a'$ for Alice, and $b$ or $b'$ for Bob. As before, we shall assume that on each run of the system, Alice can read the value of $a$ \emph{or} $a'$, but not both; and similarly, Bob can only read one of $b$ or $b'$. (In more physical terms, only one of these quantities can be measured in any given run.) We shall assume that Alice and Bob are spacelike separated, and hence they can perform their measurements or observe their variables independently of each other. By observing multiple runs of this scenario, we obtain probabilities $p(uv|xy)$ of obtaining the joint outcomes $x=u$ and $y = v$ when Alice selects the variable $x \in \{ a, a' \}$, and Bob selects $y \in \{ b, b' \}$.

Consider the following tabulation of values for these probabilities:
\begin{center}
\begin{tabular}{l|ccccc}
& $00$ & $01$ & $10$ & $11$  &  \\ \hline
$ab$ & $1/2$ & $0$ & $0$ & $1/2$ & \\
$ab'$ & $1/2$ & $0$ & $0$ & $1/2$ & \\
$a'b$ & $1/2$ & $0$ & $0$ & $1/2$ & \\
$a'b'$ & $0$ & $1/2$ & $1/2$ & $0$ & 
\end{tabular}
\end{center}
The entry in row $xy$ and column $uv$ gives the probability $p(uv|xy)$.

This is the PR-box of Popescu and Rohrlich \cite{popescu1994quantum}. As is well-known, it maximises the value of the CHSH expression:
\[ \EE(ab) + \EE(ab') + \EE(a'b) - \EE(a'b') = 4, \]
where
\[ \EE(xy) \; = \; \sum_{u,v} (-1)^{u+v} p(uv|xy) . \]
Thus it exceeds the Tsirelson bound \cite{tsirelson1980} of $2 \sqrt{2}$ for the maximum degree of correlation that can be achieved by any bipartite quantum  system of this form. It follows that \emph{no quantum system can give rise to this probabilistic model}. At the same time, it still satisfies No-Signalling, and hence is consistent with the relativistic constraints imposed by the spacelike separation of Alice and Bob.

We now consider the analogous form of ``explanation'' for this model which can be given in the same style as for our previous example.
The ontic or hidden variables will assign a definite value to each of the four possible measurements which can be made in this scenario: $a$, $a'$ $b$, $b'$. We shall use the notation $uu'vv'$ for the assignment
\[ a \mapsto u, \quad a' \mapsto u', \quad b \mapsto v, \quad b' \mapsto v' . \]
Following Mermin \cite{mermin1990quantum}, we shall call such assignments \emph{instruction sets}. In this case, there are $2^4 = 16$ instruction sets.
We assume that on each run of the system, such an instruction set is sampled according to a probability distribution on this 16-element set.
The values observed by Alice and Bob, given their choice of measurement settings, are those prescribed by the instruction set.

Now Bell's theorem tells us that no standard probability distribution on the instruction sets can give rise to the behaviour of the PR Box; while from the Tsirelson bound, we know that the correlations achieved by the PR box exceed those which can be realised by a quantum system.

However, consider the following signed measure on  instruction sets \cite[Example 5.2]{abramsky2011unified}:
\[ \begin{array}{lcr}
p(0000) & = & 1/2 \\
p(0001) & = & 0 \\
p(0010) & = & -1/2\\
p(0011) & = & 0 \\
p(0100) & = &  0\\
p(0101) & = & 0 \\
p(0110) & = & 1/2 \\
p(0111) & = & 0
\end{array}
\quad \quad
\begin{array}{lcr}
p(1000) & = & -1/2 \\
p(1001) & = & 1/2 \\
p(1010) & = & 1/2 \\
p(1011) & = & 0 \\
p(1100) & = & 0 \\
p(1101) & = & 0 \\
p(1110) & = & 0 \\
p(1111) & = & 0 \\
\end{array}
\]
We can check that this distribution reproduces exactly the probabilities of the PR box. For example:
\[ p(00|ab) = p(0000) + p(0001) + p(0100) + p(0101) = 1/2, \]
and similarly $p(uv|xy)$ can be obtained from this signed measure on the instruction sets for all $u,v,x,y$.
Since the ``observable events'' are exactly those corresponding to these  probabilities, the negative values occurring in the signed measure can never be observed. The probabilities of the outcomes for a given choice of measurement settings, corresponding to the rows of the PR box table, form well-defined standard probability distributions.

This is not an isolated result. In \cite[Theorem 5.9]{abramsky2011unified} it is shown that for a large class of probability models, including Bell scenarios with any numbers of agents, measurement settings and outcomes, and also contextuality scenarios including arbitrary Kochen-Specker configurations, the model can be realised by a signed measure on instruction sets if and only if it is No-Signalling.

But this brings us back to the question: what \emph{are} these negative probabilities?
The main purpose of the present paper
is to give an operational interpretation of negative probabilities, in a broadly frequentist setting, by means of a simulation in terms of standard probabilities. We shall postpone discussion of the conceptual status of this interpretation to the final section of the paper, after the ideas have been put in place.

The further structure of this paper is as follows. In Section~2, we shall lay out the simple ideas involved in our interpretation of negative probabilities at the level of abstract probability distributions. We shall only consider the case of discrete measures, which will be sufficient for our purposes.
In Section~3, we shall review observational scenarios and empirical models in the general setting studied in \cite{abramsky2011unified}, and the result relating signed measures and No-Signalling models. In Section~4, we shall develop an operational interpretation of hidden-variable models, including those involving negative probabilities. By virtue of the general result on hidden-variable models with signed measures, this yields a uniform scheme for simulating all No-Signalling boxes using only classical resources. Finally, Section~5 concludes with a discussion of the results, and some further directions.

\section{Probabilities and Signed Measures}

Given a set $X$, we write $\MM(X)$ for the set of (finite-support) \emph{signed probability measures} on $X$, \ie the set of maps $m : X \rarr \Real$ of finite support, and such that
\[ \sum_{x \in X} m(x) = 1 . \]
We extend measures to subsets $S \subseteq X$ by (finite) additivity:
\[ m(S) \; := \; \sum_{x \in S} m(x) . \]
We write $\prob(X)$ for the subset of $\MM(X)$ of measures valued in the non-negative reals; these are just the  probability distributions on $X$ with finite support.

These constructions also act on maps. Given a function $f : X \rarr Y$, we can define
\[ \MM(f) : \MM(X) \lrarr \MM(Y) :: m \mapsto [y \mapsto \sum_{f(x) = y} m(x) ] . \]
Thus $\MM(f)$ pushes measures on $X$ forwards along $f$ to measures on $Y$.

Since $\MM(f)$ will always map probability distributions to probability distributions, we can define $\prob(f) := \MM(f) |_{\prob(X)}$.
It can easily be checked that these assignments are functorial:
\[ \MM(g \circ f) = \MM(g) \circ \MM(f), \quad \MM(\id_{X}) = \id_{\MM(X)}, \]
and similarly for $\prob$.

Now we come to the basic idea of our approach, which is to interpret signed measures by ``pushing the minus signs inwards''. That is, we take basic events to carry an additional bit of information, a sign or ``probability charge''. Moreover, occurrences of the same underlying event of opposite sign cancel. In this fashion, negative probabilities arise from standard probabilities on signed events.\footnote{This intuitive way of looking at signed measures has, \textit{grosso modo},  appeared in the literature, e.g.~in a paper by Burgin \cite{burgin2010interpretations}. However, the details in \cite{burgin2010interpretations} are very different to our approach. In particular, the notion of signed relative frequencies  in \cite{burgin2010interpretations}, which is defined as a difference of ratios rather than a ratio of differences, is not suitable for our purposes.}

More formally, given a set $X$, we take the signed version of $X$ to be the disjoint union of two copies of $X$, which we can write as
\[ \Xpm \; := \; \{ (x, \vs) \mid x \in X, \vs \in \{ {+}, {-} \} \} . \]
Also, given a map $f : X \rarr Y$, we can define a map $f^{\pm} : \Xpm \rarr \Ypm$ by
\[ f^{\pm} : (x, \vs) \mapsto (f(x), \vs), \quad \vs \in \{ {+}, {-} \} . \]
We shall use the notation $\xp, \xm$ rather than $(x, {+})$, $(x, {-})$.
Given $S \subseteq X$, we shall write 
\[ S^{+} := \{ x^{+} \mid x \in S \},  \;\; S^{-} := \{ x^{-} \mid x \in S \} \; \subseteq \; X^{\pm} . \]

The representation of a signed measure on a sample space $X$ by a probability measure on $\Xpm$ is formalised by the map
\[ \theta_X : \MM(X) \lrarr \prob(\Xpm) \]
given by
\[ \theta_X(m)(\xp) = \left\{ \begin{array}{ll} m(x) / K, & m(x) > 0 \\
0, & \mbox{otherwise}
\end{array}
\right.
\]
\[ \theta_X(m)(\xm) = \left\{ \begin{array}{ll} | m(x) | / K, & m(x) < 0 \\
0, & \mbox{otherwise}
\end{array}
\right.
\]
where $K = \sum_{x \in X} | m(x) |$ is a normalisation constant.

Note that the probability measures in the image of $\theta_X$ have some special properties.
In particular, if we define $W(d)$, for $d \in \prob(\Xpm)$, by
\[ W(d) \; := \; d(X^{+}) - d(X^{-}), \]
then if $d = \theta_X(m)$, we have
\[ W(d) \; = \; \sum_{x \in X} m(x)/K  \; = \; 1/K \; > \; 0 . \] 
We write $\Pp(X) \, :=\, \{ d \in \prob(\Xpm) \mid W(d) > 0\}$. Thus $\theta_X$ cuts down to a map
\[ \theta_X : \MM(X) \lrarr \Pp(X) . \]
Note also that for any map $f : X \rarr Y$, and $d \in \prob(\Xpm)$:
\[ W(\prob(\fpm)(d)) = W(d) . \]
Hence we can extend $\Pp$ to a functor by $\Pp(f) := \prob(\fpm)$.

We can recover a signed measure on $X$ from a probability distribution in $\Pp(X)$ by an inverse process to $\theta_X$.
Formally, this is given by a map
\[ \eta_X : \Pp(X) \lrarr \MM(X) :: d \mapsto [ x \mapsto (d(\xp) - d(\xm)) / W(d) ] . \]
This map incorporates the idea that positive and negative occurrences of a given event cancel.

The following result will be used in showing the correctness of our simulation scheme in Section~\ref{signsec}.

\begin{proposition}
\label{corrprop}
The following diagram commutes, for all sets $X$ and $Y$ and functions $f : X \rarr Y$:
\begin{diagram}[4em]
\MM(X) & \rTo^{\MM(f)} & \MM(Y) \\
\dTo^{\theta_X} & & \uTo_{\eta_Y} \\
\Pp(X) & \rTo_{\Pp(f)} & \Pp(Y) \\
\end{diagram}
\end{proposition}
\begin{proof}
We write $d := \theta_X(m)$, and calculate pointwise on $y \in Y$:
\begin{align*}
\eta_Y \circ \Pp(f) \circ d (y) 
&\;= \;\frac{d((\fpm)^{-1}(y^{+})) \; - \; d((\fpm)^{-1}(y^{-}))}{d(X^{+}) \; - \; d(X^{-})} \\
&\;= \;\sum_{f(x)=y, m(x)>0} m(x) \; - \; \sum_{f(x)=y, m(x)<0} |m(x)|  \\
&\;= \;\sum_{f(x)=y} m(x) \\
&\;= \;\MM(f)(m)(y) . \\
\end{align*}
\end{proof}

Intuitively, this says that pushing a signed measure $m$ forwards along $f$ can be performed by simulating the measure by a probability distribution $d = \theta_X(m)$ on signed events, pushing $d$ forwards along $f$,   and then interpreting the resulting probability distribution back as a signed measure via the map $\eta_Y$.

\section{Observational Scenarios and Empirical Models}

An \emph{observational scenario} is a structure $(X, \UU, O)$, where:
\begin{itemize}
\item $X$ is a set of measurements.
\item $\UU$ is a family of non-empty subsets of $X$ with $\bigcup \UU = X$, representing the compatible sets of measurements --- those which can be  performed together.
\item $O$ is a set of measurement outcomes.
\end{itemize}

For example, in the scenario for the PR box described in the Introduction, we have:
\begin{itemize}
\item $X = \{ a, a', b, b' \}$.
\item The measurement contexts are the choice of measurement settings by Alice and Bob:
\[ \{ a, b \}, \{ a', b \}, \{ a, b' \}, \{ a', b' \} . \]
\item The outcomes are $O = \{ 0, 1 \}$.
\end{itemize}

An empirical model for such a scenario is a family of probability distributions $\{ d_U \}_{U \in \UU}$, with $d_U \in \prob(O^U)$.
Here $O^U$ is the set of all functions $s : U \rarr O$. Such functions represent basic events in the measurement context $U$, where the measurements in $U$ are performed, and the outcome $s(x)$ is observed for each $x \in U$.

In the case of the PR box, the distributions correspond to the rows of the table, indexed by the measurement contexts $xy$.

We define an operation of restriction on signed measures, which is a general form of marginalization: if $U \subseteq V$ and $m \in \MM(O^V)$, then $m |_U \in \MM(O^U)$ is defined by:
\[ m |_U(s) = \sum_{t |_U = s} m(t) . \]
Note that  $m |_U = \MM(\rho^V_U)(m)$, where 
\[ \rho^V_U : O^V \lrarr O^U :: s \mapsto s |_U . \]
Also, if $d \in \prob(O^V)$, then $d |_U \in \prob(O^U)$.

A model $\{ d_U \}_{U \in \UU}$ is \emph{compatible} if for all $U, V \in \UU$:
\[ d_U |_{U \cap V} \; = \; d_V |_{U \cap V} . \]
As shown in \cite{abramsky2011unified}, compatibility can be seen as a general form of \emph{no-signalling}.

A \emph{global section} for an empirical model $\{ d_U \}_{U \in \UU}$ is a distribution $d \in \prob(O^X)$ such that, for all $U \in \UU$:
\[ d |_U = d_U . \]
As shown in \cite{abramsky2011unified}, a global section can be seen as a canonical form of local hidden variable model.

A \emph{signed global section} is a signed measure $m \in \MM(O^X)$ such that $m |_U = d_U$ for all $U \in \UU$.
We have the following result from \cite{abramsky2011unified}.

\begin{theorem}
\label{npnsthm}
An empirical model $\{ d_U \}_{U \in \UU}$ is no-signalling if and only if it has a signed global section.
\end{theorem}

\section{Operational interpretation of hidden-variable models}

We begin with standard local hidden-variable models. We shall follow the expository scheme due to Mermin phrased in terms of \emph{instruction sets} \cite{mermin1990quantum}, which is encapsulated in the diagram in Figure~1.

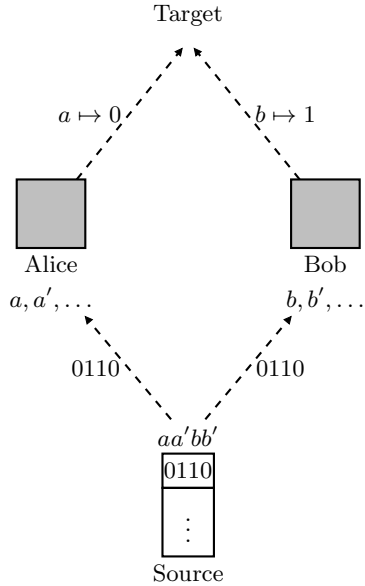
\begin{figure}
\label{merfig}
\begin{center}
{\small
\input{mermin.tex}
}
\end{center}
\caption{The Mermin instruction set picture}
\end{figure}
Here the agents or experimenters Alice and Bob each receive a stream of particles; for each particle, they choose a measurement setting, and observe an outcome. The stream of observed joint outcomes is collected at a target, and provides the statistical data on which the empirical model is based.

In this way of visualising hidden variables, the pairs of particles are generated by some source. Reflecting the idea of local realism, each particle comes  with an ``instruction set'' which specifies an outcome for every possible measurement which can be performed. In the pictured scenario, where we are considering measurement settings $a$, $a'$ for Alice and $b$, $b'$ for Bob, such an instruction set will specify an outcome, $0$ or $1$, for each of these four measurements, and hence is depicted as a string of four bits, which correspond to outcomes for $a, a', b, b'$.
When such a particle arrives e.g. at Alice and she performs the measurement $a$, the instruction set $0110$ dictates that she will observe the outcome $0$; while if Bob performs the measurement $b$ on a particle with the same instruction set he will observe the outcome $1$.
To account for the fact that the same measurements may yield different outcomes, we depict the source as obtaining the stream of particles with inscribed instruction sets by sampling some probability distribution on the space of instruction sets.

The content of Bell's theorem and related no-go results is that there is \emph{no probability distribution on instruction sets} which can account for the probabilistic behaviour which is predicted by quantum mechanics, and highly confirmed by experiment.

In mathematical terms, instruction sets are just functions in $O^X$; and the non-existence of a probability distribution on instruction sets which recovers the observed behaviour is exactly the non-existence of a global section for the corresponding empirical model.

\subsection{Formalisation of the instruction set picture}

We now give a  formal account of this standard picture. 

Firstly, we give a more explicit, frequentist description of the operational reading we have in mind.
\begin{enumerate}
\item We fix some probability distribution $d$ on instruction sets.
\item The \emph{source} produces a stream of instruction sets $t_n$ by sampling repeatedly according to $d$. 
It sends a stream of pairs of particles inscribed with the corresponding instruction sets $t_n$ to Alice and Bob.
\item \emph{Alice} and \emph{Bob} act independently
of the source.  Alice receives a stream of particles from the source, chooses and performs  a corresponding stream of local measurements $a_n$, and sends her measurement choices $a_n$ and the outcomes $u_n$ to the target.  Similarly, Bob receives a stream of particles, chooses measurements $b_n$, and sends his measurement choices $b_n$ and the outcomes $v_n$ to the target. Note that the joint outcome $u_nv_n$ of a measurement context $a_{n}b_{n}$ specified by an instruction set $t_n$ is given by $u_n = t_n(a_n)$, $v_n = t_n(b_n)$.
\item The \emph{target} receives the streams of measurement choices and outcomes from Alice and Bob.
It combines these into a stream $\tau$ of joint outcomes, with $\tau_n = u_{n}v_n$. For each choice of measurement context $xy$ and joint outcome $uv$, and for each $n$, it can compute the relative frequency $r_n(uv|xy)$ of $uv$ in the initial segment of $\tau_{xy}$ of length $n$.
Here $\tau_{xy}$ is the restriction of $\tau$ to the subsequence of those $\tau_i$ such that $a_i = x$ and $b_i = y$.
\item The limiting value of these relative frequencies $r_n(uv|xy)$ is taken to be the probability $p(uv | xy)$.
\end{enumerate}

We comment on a number of points raised by this description.

\begin{itemize}
\item Firstly, note that the r\^ole of the various agents in this protocol is clearly delineated by the information flows it makes explicit.
\begin{itemize}
\item Alice and Bob cannot predict the outcome, since they must accept unknown particles from the source.
\item Although the instruction sets generated by the source determine the outcomes \emph{given the choice of measurements}, the source cannot predict which measurements Alice and Bob will select.
\end{itemize}
These informational independence notions are reflected in the standard assumption of independence between the distribution governing the instruction sets, and the choices made by Alice and Bob. This is usually referred to as Free Choice of Measurements, or $\lambda$-independence \cite{dickson1999quantum}.
Without this assumption, the protocol trivialises, and arbitrary behaviour can be generated \cite{abramsky2013no}.
\item The target is usually not made explicit. However, since Alice and Bob are assumed to be spacelike separated, in order for it to be possible to obtain empirical data on the correlations between their outcomes, it is necessary to assume that their future light-cones intersect.
\end{itemize}

We now formalise this informal frequentist account in terms of standard probability theory.

Firstly, we recall standard notation for indicator functions. If $U \subseteq X$, we write $\IF_U : X \rarr \Real$ for the function
\[ \IF_U : x \mapsto \left\{ \begin{array}{ll}
1, & x \in U \\
0, & x \not\in U.
\end{array} \right.
\]
If we fix a distribution $d \in \prob(X)$, we can regard $\IF_U$ as a random variable with respect to $d$.
Note that, writing $\EE(R)$ for the expectation of a random variable $R$:
\[ \EE(\IF_U) \; = \; d(U) . \]

We write $X := \{ a, a', b, b' \}$, $O := \{ 0, 1 \}$, and $\II := O^X$ for the set of instruction sets.
We write $X_A := \{ a, a' \}$ for the set of Alice's measurement choices, and similarly $X_B := \{ b, b' \}$.

We are given a probability distribution $d \in \prob(\II)$.
We also assume a probability distribution $d_{AB}$ on measurement choices $xy$, which is assumed to be independent of $d$, reflecting our assumption that Alice and Bob's measurement choices are independent of the source. We shall assume that $d_{AB}(xy) > 0$ for all measurement contexts $xy$, \ie that all measurements have some chance of being performed. If this were not the case, we could simply exclude measurements which could never be performed from the scenario.

Thus we have a probability distribution $\mu$ on $\II \times X_A \times X_B$:
\[ \mu(t, x, y) \; = \; d(t) d_{AB}(xy) . \]

The stream of data at the target comprises items of the form $uvxy \in O \times O \times X_A \times X_B$.
This is determined by the instruction set generated at the source and the measurement choices made by Alice and Bob, as specified by the function 
\[ f : \II \times X_A \times X_B \lrarr O \times O \times X_A \times X_B \; :: \; (t, x, y) \mapsto (t(x), t(y), x, y) . \]
We can use the functorial action of $\prob$ to push $\mu$ forward along $f$ to yield a probability distribution $\nu := \prob(f)(\mu)$ on $O \times O \times X_A \times X_B$. This can be defined explicitly as follows.
Given $u$, $v$, $x$, $y$, we define $U(uvxy) \subseteq \II$:
\[ U(uvxy) \; := \; \{ t \in \II \mid t(x) = u \AND t(y) = v \} . \]
Now $\nu(uv,xy) = d(U(uvxy))d_{AB}(xy)$.
Note that $d(U(uvxy)) = \prob(\rho^{\II}_{xy})(d) (uv)$.

The conditional probability for the target to observe outcomes $uv$ given measurement choices $xy$ is:
\[ \nu(uv|xy) \; = \; \frac{d(U(uvxy))d_{AB}(xy)}{d_{AB}(xy)} \; = \; d(U(uvxy)) . \]
Thus the stochastic process at the target for observing outcomes $uv$ given measurement settings $xy$ is modelled by the i.i.d. sequence
of random variables $\XX_n$, where for all $n$, $\XX_n = \IF_{U(uvxy)}$.
The relative frequencies observed at the target are represented by the means of these random variables:
\[  r_n(uv|xy) \; = \; \frac{1}{n} \sum_{i=1}^n \XX_i . \]
Using the Strong Law of Large Numbers \cite{billingsley2008probability}, we can calculate that
\[ p(uv|xy) \; = \; d |_{xy} (uv) \; = \; \prob(\rho^{\II}_{xy})(d) (uv) \; = \; d(U(uvxy)) \; = \; \EE(\IF_{U(uvxy)}) \; =_{\mbox{a.e.}} \; \lim_{n \rarr \infty} \frac{1}{n} \sum_{i=1}^n \XX_i . \]
This provides a precise statement of the agreement of the operational protocol with the abstract formulation.

\subsection{Signed probabilities and no-signalling models}
\label{signsec}

We now return to Theorem~\ref{npnsthm} and use our account of negative probabilities to give it an operational interpretation, which we formulate as a refinement of the Mermin picture.

We use the signed version of the Mermin instruction set scenario depicted in Figure~2.

\begin{figure}
\begin{center}
{\small
\input{signedmermin.tex}
}
\end{center}
\caption{Signed instruction sets}
\end{figure}
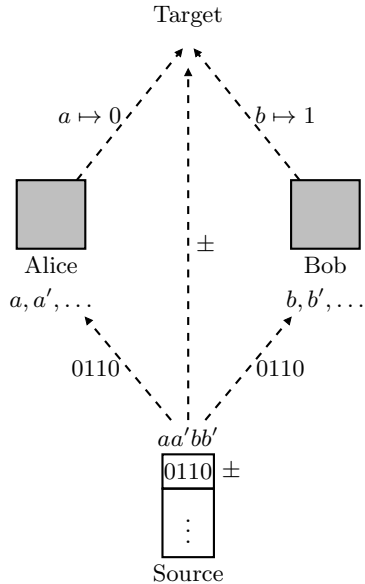

We have the same picture as before, subject to the following refinements:
\begin{itemize}
\item The particles come with an additional bit of information in addition to the instruction set: a \emph{sign}.
\item The source repeatedly samples these signed instruction sets according to a (standard) probability distribution, and sends the particles with their instruction sets to Alice and Bob, who choose their measurements and observe outcomes as before.
\item The joint outcomes are collected at the target, which also receives the information concerning the signs of the particles.
\item The target uses the  signs to compute \emph{signed relative frequencies} on the stream of joint outcomes, and hence to recover a signed measure on the joint outcome. In certain cases, this signed measure may in fact be a \textit{bona fide} probability measure.

The signed relative frequencies incorporate the idea of cancelling positive and negative occurrences of events. The signed relative frequency of an \emph{unsigned} event in a \emph{signed} ensemble is the difference between its number of positive and negative occurrences in the ensemble, normalised by the total weight of the ensemble, which is the difference between the total numbers of positive and negative occurrences in the ensemble.
\end{itemize}

We shall now set out the simulation scheme in more precise terms. The analysis is very similar to the unsigned case, the key difference being the use of signed relative frequencies.

We are given a no-signalling empirical model with probabilities $p(uv|xy)$.

\begin{enumerate}
\item By Theorem~\ref{npnsthm}, we can find  a signed measure $m$ on the instruction sets which yields the observed probabilities by marginalization: $m |_{xy}(uv) = p(uv|xy)$ for all $u,v,x,y$.
\item We form the probability distribution $d = \theta_{\II}(m)$ on signed instruction sets.
\item The \emph{source} produces a stream of signed instruction sets $\spm_n$ by repeatedly sampling from $d$. 
It sends a stream of particles inscribed with the corresponding instruction sets $s_n$ to Alice and Bob, and sends the signs  to the target.
\item \emph{Alice} and \emph{Bob} act independently of the scheme, in exactly the same manner as in the unsigned case.
They send their measurement choices $a_n$, $b_n$, and the corresponding outcomes $u_n$, $v_n$, where $u_n = s_n(a_n)$ and $v_n = s_n(b_n)$, to the target.
\item The \emph{target} receives the streams of outcomes and measurement choices from Alice and Bob, and the stream of signs from the source.
It uses these to compute the signed relative frequencies $\rpm_n(uv|xy)$ for joint outcomes $uv$ given measurement choices $xy$.
\item The limiting value of these signed relative frequencies $\rpm_n(uv|xy)$ is taken to be the probability $p(uv | xy)$.
\end{enumerate}

The stream of data received at the target comprises items of the form 
\[ (uv)^{\vs}xy \in (O \times O)^{\pm} \times X_A \times X_B , \qquad \vs \in \{ +, - \} . \]
This is determined by the signed instruction set generated at the source, and the measurement choices made by Alice and Bob, as specified by the function
\[ g : \II^{\pm} \times X_A \times X_B \lrarr (O \times O)^{\pm} \times X_A \times X_B \; :: \; (s^{\vs}, x, y) \mapsto ((s(x), s(y))^{\vs}, x, y) . \]
We can use this function to push forward the measure $\mupm$ on $\II^{\pm} \times X_A \times X_B $, defined by
\[ \mupm(s^{\vs}, x, y) \; := \; d(\spm) d_{AB}(xy) \]
to $\nupm := \Pp(g)(\mupm)$ on $(O \times O)^{\pm} \times X_A \times X_B$. This measure is defined explicitly by:
\[ \nupm((uv)^{\vs}xy) \; = \; d(U^{\vs}(uvxy)) d_{AB}(xy) . \]

As in the unsigned case, because of the product form of this measure, which corresponds to the independence of the measurement choices from the source distribution, conditioning on measurement choices $xy$ leads to the probability $d(U^{\vs}(uvxy))$ for signed outcomes $(uv)^{\vs}$.

We define the following i.i.d. sequences of random variables:
\[ \rv{P}^{+}_n \; := \; \IF_{U^{+}(uvxy)}, \quad \rv{P}^{-}_n \; := \; \IF_{U^{-}(uvxy)}, \quad \rv{Q}^{+}_n \; := \; \IF_{\Ixy^{+}}, \quad \rv{Q}^{-}_n \; := \; \IF_{\Ixy^{-}} . \]
The process of forming signed relative frequencies $\rpm_n(uv|xy)$ at the target is modelled by the sequence of random variables $\rv{S}_n$, where:
\[ \rv{S}_n \; := \; \frac{\sum_{i=1}^{n}  \rv{P}^{+}_n \; - \; \sum_{i=1}^{n} \rv{P}^{-}_n}{\sum_{i=1}^{n} \rv{Q}^{+}_n \; - \; \sum_{i=1}^{n} \rv{Q}^{-}_n} . \]
The correctness of our simulation is now expressed by the following result.

\begin{theorem}
For all $u,v,x,y$:
\[ \lim_{n \rarr \infty} \rv{S}_n \; =_{\mbox{a.e.}} \; p(uv|xy) . \]
\end{theorem}
\begin{proof}
By the Strong Law of Large Numbers, 
\[ \lim_{n \rarr \infty} \frac{1}{n} \sum_{i=1}^n \rv{P}^{+}_i  \; =_{\mbox{a.e.}} \; \EE(\IF_{U^{+}(uvxy)}) \; = \; d(U^{+}(uvxy)) . \]
Unpacking this more carefully (see e.g. \cite[Theorem 6.1, p.85]{billingsley2008probability}), the random variables $\rv{P}_n$ act on the probability space $\Seq = (\II^{\pm})^{\omega}$, the product of countably many copies of $\II^{\pm}$, with product measure $d^{\omega}$. The action is given by:
\[ \rv{P}_n \; = \; \IF_{U^{+}(uvxy)} \circ \pi_n . \]
The  Strong Law asserts that, for some set $Z_1$ of measure zero in $\Seq$, for all $\sigma \in \Seq \setminus Z_1$:
\[ \lim_{n \rarr \infty}  \frac{1}{n} \sum_{i=1}^n \rv{P}^{+}_i (\sigma) \; = \; d(U^{+}(uvxy)) . \]
Similarly, outside sets $Z_2$, $Z_3$, $Z_4$ of measure 0, we have
\begin{align*}
\lim_{n \rarr \infty} \frac{1}{n} \sum_{i=1}^n \rv{P}^{-}_i  & =  d(U^{-}(uvxy)) \\
\lim_{n \rarr \infty} \frac{1}{n} \sum_{i=1}^n \rv{Q}^{+}_i  & =  d(\Ixy^{+}) \\
\lim_{n \rarr \infty} \frac{1}{n} \sum_{i=1}^n \rv{Q}^{-}_i  & =  d(\Ixy^{-}).
\end{align*}
Since $d = \theta_{\II}(m)$, 
\[ d(\Ixy^{+}) - d(\Ixy^{-}) \; = \; W(d) \; > \; 0 , \]
and hence for all $\sigma \in \Seq \setminus (Z_3 \cup Z_4)$, for all but finitely many $n$:
\[ \frac{1}{n} \sum_{i=1}^n \rv{Q}^{+}_i (\sigma) \; - \; \frac{1}{n} \sum_{i=1}^n \rv{Q}^{-}_i (\sigma) \; > \; 0 . \]
Now $Z := Z_1 \cup Z_2 \cup Z_3 \cup Z_4$ has measure 0, and for all $\sigma \in \Seq \setminus Z$, by standard pointwise properties of limits:
\begin{align*}
\lim_{n \rarr \infty} \rv{S}_n (\sigma) & \; = \; \lim_{n \rarr \infty} \frac{\frac{1}{n} (  \sum_{i=1}^{n} \rv{P}^{+}_i (\sigma) \; - \;  \sum_{i=1}^{n} \rv{P}^{-}_i (\sigma))}{\frac{1}{n} ( \sum_{i=1}^{n} \rv{Q}^{+}_i (\sigma) \; - \;  \sum_{i=1}^{n} \rv{Q}^{-}_i (\sigma))} \\
& \; = \; \frac{d(U^{+}(uvxy)) \; - \; d(U^{-}(uvxy))}{d(\Ixy^{+}) \; - \; d(\Ixy^{-})} \\
& \;= \; \eta(\Pp(\rho^{\II}_{xy})(\theta_X(m)))(uv) \\
& \;= \; \MM(\rho^{\II}_{xy})(m) (uv)  \qquad \qquad \qquad \mbox{by Proposition~\ref{corrprop}} \\
& \; = \; m |_{xy} (uv) \\
& \; = \; p(uv|xy) .
\end{align*}
\end{proof}

\section{Discussion}

A first point to  make is that the scheme we described in the previous section was formulated for systems of type $(2,2,2)$; that is, with two agents, two measurements per agent, and two outcomes per measurement. This was to avoid notational complications.
It is clear that the same scheme would apply to Bell-type systems with any numbers of agents, measurement settings and outcomes.
It is less clear how to proceed with other kinds of contextuality scenarios, although the result in Theorem~\ref{npnsthm} certainly applies to such scenarios.

The interpretation we have given of negative  probabilities is operational in nature.  It can be implemented in a physical scheme as summarized in the signed instruction set diagram in Figure 2.  However, one should think of this scheme as a \emph{simulation}, rather than  a direct description of a fundamental physical process. The fact that it applies to arbitrary no-signalling systems, including superquantum devices such as PR boxes, which are generally believed to go beyond what is physically realizable, compels caution in this respect.

At the same time, the nature of the simulation, which respects relativistic constraints and uses only classical probabilistic devices, provides interesting food for thought. After all, this is a concrete way of thinking about entanglement, and even superquantum correlations, in terms of familiar-seeming devices: one can e.g.~think of the source as generating its stream of signed particles by drawing coloured billiard balls from an urn. The subsequent passages of the particles are entirely classical. The only non-standard element of the process is the cancellation of positive and negative events effected by forming the signed relative frequencies. Can one find some structural features within this mode of description of non-local correlations which can help to delineate the quantum/superquantum boundary?

Among the features which it may be interesting to study from this point of view are the rates of convergence of the stochastic processes described in the previous section. If cancellation of positive events by negative ones can occur with unbounded delays, there may be some form of \emph{retrocausality} hidden in the computation of the signed relative frequencies. Do quantum processes admit bounds on cancellation which ensure that causal anomalies do not arise? It may also be interesting to compare the entropies of the simulating (unsigned) and simulated (signed) processes. Computational efficiency may also provide a useful perspective.

While we certainly do not claim to have solved any mysteries, we hope to have provided a novel way of thinking about some of the mysterious features of the quantum --- and even the superquantum --- world. 

\bibliographystyle{plain}
\bibliography{bdbib}

\end{document}

%% file: mermin.tex
\begin{tikzpicture}[scale=2.54]
\ifx\dpiclw\undefined\newdimen\dpiclw\fi
\global\def\dpicdraw{\draw[line width=\dpiclw]}
\global\def\dpicstop{;}
\dpiclw=0.8bp
\dpicdraw[fill=lightgray](0,-0.17951) rectangle (0.359021,0.17951)\dpicstop
\dpicdraw[fill=lightgray](1.436083,-0.17951) rectangle (1.795104,0.17951)\dpicstop
\draw (0.17951,-0.17951) node[below=-0.269266bp]{Alice};
\draw (1.615594,-0.17951) node[below=-0.269266bp]{Bob};
\draw (0.17951,-0.448776) node{$a, a', \ldots$};
\draw (1.615594,-0.448776) node{$b, b', \ldots$};
\dpicdraw (0.762919,-1.436083) rectangle (1.032185,-1.256573)\dpicstop
\draw (0.897552,-1.346328) node{$0110$};
\dpicdraw (0.762919,-1.795104) rectangle (1.032185,-1.436083)\dpicstop
\draw (0.897552,-1.615594) node{$\vdots$};
\draw (0.897552,-1.256573) node[above=-0.269266bp]{$aa'bb'$};
\draw (0.897552,-1.795104) node[below=-0.269266bp]{Source};
\dpicdraw[dashed](0.807797,-1.077063)
 --(0.382005,-0.566112)\dpicstop
\filldraw[line width=0bp](0.395795,-0.55462)
 --(0.359021,-0.538531)
 --(0.368214,-0.577604) --cycle
\dpicstop
\draw (0.584123,-0.808653) node[left=-0.269266bp]{$0110$};
\dpicdraw[dashed](0.987307,-1.077063)
 --(1.413099,-0.566112)\dpicstop
\filldraw[line width=0bp](1.42689,-0.577604)
 --(1.436083,-0.538531)
 --(1.399309,-0.55462) --cycle
\dpicstop
\draw (1.210982,-0.808653) node[right=-0.269266bp]{$0110$};
\draw (0.897552,0.94243) node[above=-0.269266bp]{Target};
\dpicdraw[dashed](0.314143,0.17951)
 --(0.843326,0.837836)\dpicstop
\filldraw[line width=0bp](0.857317,0.82659)
 --(0.865819,0.865819)
 --(0.829334,0.849083) --cycle
\dpicstop
\draw (0.589282,0.521796) node[left=-0.269266bp]{$a \mapsto 0$};
\dpicdraw[dashed](1.480961,0.17951)
 --(0.951779,0.837836)\dpicstop
\filldraw[line width=0bp](0.96577,0.849083)
 --(0.929285,0.865819)
 --(0.937787,0.82659) --cycle
\dpicstop
\draw (1.205822,0.521796) node[right=-0.269266bp]{$b \mapsto 1$};
\end{tikzpicture}

%% file: signedmermin.tex
\begin{tikzpicture}[scale=2.54]
\ifx\dpiclw\undefined\newdimen\dpiclw\fi
\global\def\dpicdraw{\draw[line width=\dpiclw]}
\global\def\dpicstop{;}
\dpiclw=0.8bp
\dpicdraw[fill=lightgray](0,-0.17951) rectangle (0.359021,0.17951)\dpicstop
\dpicdraw[fill=lightgray](1.436083,-0.17951) rectangle (1.795104,0.17951)\dpicstop
\draw (0.17951,-0.17951) node[below=-0.269266bp]{Alice};
\draw (1.615594,-0.17951) node[below=-0.269266bp]{Bob};
\draw (0.17951,-0.448776) node{$a, a', \ldots$};
\draw (1.615594,-0.448776) node{$b, b', \ldots$};
\dpicdraw (0.762919,-1.436083) rectangle (1.032185,-1.256573)\dpicstop
\draw (0.897552,-1.346328) node{$0110$};
\dpicdraw (0.762919,-1.795104) rectangle (1.032185,-1.436083)\dpicstop
\draw (0.897552,-1.615594) node{$\vdots$};
\draw (0.897552,-1.256573) node[above=-0.269266bp]{$aa'bb'$};
\draw (0.897552,-1.795104) node[below=-0.269266bp]{Source};
\draw (1.032185,-1.346328) node[right=-0.269266bp]{$\plusminus$};
\dpicdraw[dashed](0.807797,-1.077063)
 --(0.382005,-0.566112)\dpicstop
\filldraw[line width=0bp](0.395795,-0.55462)
 --(0.359021,-0.538531)
 --(0.368214,-0.577604) --cycle
\dpicstop
\draw (0.584123,-0.808653) node[left=-0.269266bp]{$0110$};
\dpicdraw[dashed](0.987307,-1.077063)
 --(1.413099,-0.566112)\dpicstop
\filldraw[line width=0bp](1.42689,-0.577604)
 --(1.436083,-0.538531)
 --(1.399309,-0.55462) --cycle
\dpicstop
\draw (1.210982,-0.808653) node[right=-0.269266bp]{$0110$};
\draw (0.897552,0.94243) node[above=-0.269266bp]{Target};
\dpicdraw[dashed](0.314143,0.17951)
 --(0.843326,0.837836)\dpicstop
\filldraw[line width=0bp](0.857317,0.82659)
 --(0.865819,0.865819)
 --(0.829334,0.849083) --cycle
\dpicstop
\draw (0.589282,0.521796) node[left=-0.269266bp]{$a \mapsto 0$};
\dpicdraw[dashed](1.480961,0.17951)
 --(0.951779,0.837836)\dpicstop
\filldraw[line width=0bp](0.96577,0.849083)
 --(0.929285,0.865819)
 --(0.937787,0.82659) --cycle
\dpicstop
\draw (1.205822,0.521796) node[right=-0.269266bp]{$b \mapsto 1$};
\dpicdraw[dashed](0.897552,-1.077063)
 --(0.897552,0.727017)\dpicstop
\filldraw[line width=0bp](0.915503,0.727017)
 --(0.897552,0.762919)
 --(0.879601,0.727017) --cycle
\dpicstop
\draw (0.897552,-0.158187) node[right=-0.269266bp]{$\plusminus$};
\end{tikzpicture}